\documentstyle[preprint,aps]{revtex}
\tightenlines
\def\be{\begin{equation}}
\def\ee{\end{equation}}
\def\bea{\begin{eqnarray}}
\def\eea{\end{eqnarray}}
\def\ba{\begin{array}}
\def\ea{\end{array}}
\def\a{\alpha}

\def\d{\delta}

\def\0{$\Gamma_0$}
\def\l{\lambda}

\def\s{\sigma}
\def\L{${\cal L}$ }

\begin{document}
\draft

\title{Self-dual property of  the  Potts model in one dimension}
  
\author{F. Y. Wu \\
Department of Physics\\
 Northeastern University, Boston,
Massachusetts 02115}

\maketitle

\begin{abstract}
A duality relation is derived for the Potts model in one dimension
from a graphical consideration.
It is shown that the partition function is self-dual with
 the  nearest-neighbor interaction
and the external field  appearing as dual parameters.
Zeroes of the partition function are analyzed.  Particularly, we  show that
the duality relation 
implies a circle theorem in the complex temperature
plane for the one-dimensional  Ising model.
  \end{abstract}

\vskip 1cm

\section{Introduction}
It is  now well-known that the  Potts model 
\cite{potts,wuPotts} possesses
a duality relation in two dimensions. 
 In fact, it is using this duality relation that Potts \cite{potts}
first determined the  transition temperature 
for the square lattice.  The purpose of this note is to point out 
a self-dual relation of 
the Potts model in one dimension,
  a curious and somewhat surprising result
 which appears to have escaped heretofore attention.
For the Ising model this duality 
  implies a new circle theorem.

Consider  $N$ Potts spins placed on a ring interacting 
with nearest-neighbor interactions $K$ and  an external field $L$
for one specific spin state. 
Number the sites by $i=1,2,\cdots,N$ and denote the spin state at site $i$
by $\s_i=1,2,\cdots,q$.  
The partition function is 
\be
Z_N(q;K, L) = \sum_{\s_i= 1}^q \prod_{i=1}^N T(\s_i, \s_{i+1}) \label{partition}
\ee
with $\s_{N+1} = \s_1$ and
\be
T(\s_i, \s_{i+1}) = {\rm exp}[K\d_{\rm Kr}(\s_i,\s_{i+1})] \>
{\rm exp}[L\d_{\rm Kr}(\s_i,1)], \label{tmatrix}
\ee
where $\d_{\rm Kr}$ is the Kronecker delta function. 
 In ensuing discussions we shall make use of
the  fact that, due to symmetry, the partition function (\ref{partition})
is the same if  the external field is applied to any 
spin state, namely, if
$\d_{\rm Kr}(\s_i,1)$  in (\ref{tmatrix}) is replaced by $\d_{\rm Kr}(\s_i,\a)$
 for any $\a=2,3,\cdots,q$.

Regarding $T(\s,\s')$ as  the elements of a
$q\times q$  matrix ${\bf T}$,
then the partition function (\ref{partition})
can be evaluated by the standard technique of the transfer matrix.  
The characteristic equation for ${\bf T}$ is
\be
{\det} \,\left|\,\matrix {e^{K+L}-\l & e^L & \ldots & e^L \cr
                    1      & e^K-\l & \ldots &  1   \cr
                   \vdots & \vdots&\ddots & \vdots \cr
                    1      & 1    & \ldots &  e^K-\l  \cr}\,\right|=0.\label{char}
\ee
This equation can be factorized   \cite{eigen} leading to eigenvalues
  $e^K-1$, which is $(q-2)$-fold degenerate,
and $\l_\pm$,  which are the two roots of the quadratic equation
\be
\l^2 -(e^{K+L} + e^K +q-2) \l +e^L(e^K-1)(e^K+q-1) =0. \label{lambda}
\ee
This gives rise to the following explicit expression for the partition function, 
\be
Z_N(q;K, L) =  \l_+^N +\l_-^N +(q-2)(e^K-1)^N . \label{partition1}
\ee
Explicitly, we have
\be
\l_\pm ={1\over 2} \biggl[ e^{K+L} +e^K+q-2
\pm \biggl((e^{K+L} - e^K -q+2)^2 +4(q-1)e^L\biggr)^{1/2}\>\biggr].  \label{root}
\ee
Note that terms involving branch cuts
  are cancelled in expanding (\ref{partition1}), and as a 
consequence  $Z_N(q;K, L)$ is indeed a polynomial in $e^K$ and $e^L$. 

\section{The duality relation}
Our main result is the  self-dual relation for the partition function
(\ref{partition1})
\be
[(e^K-1)(e^L-1)]^{-N/2}\>{Z_N(q;K,L)} =[(e^{K^*}-1)
(e^{L^*}-1)]^{-N/2}\>{Z_N(q;K^*,L^*)} ,
\label{duality}
\ee
where the dual variables $K^*$ and $L^*$ are related to
$L$ and $K$, respectively, by
\be
(e^{K^*}-1) (e^L-1) = q, \hskip 1cm
(e^{L^*}-1) (e^K-1) = q. \label{dualvariable}
\ee
The validity of   (\ref{duality}) 
can be explicitly verified using (\ref{partition1}) for $q=2$ and for specific values of $N=2,3,\cdots$.
 But its  validity for 
general $q$ and $N$  is not very
obvious by looking at the solution.
 
To establish (\ref{duality}) for general $q$ and $N$, 
we construct  a lattice ${\cal L}$ of $N+1$ sites as shown in Fig. 1,
by introducing an extra (ghost) spin 
 interacting with all $N$ sites with an equal interaction $L$.
Note that the lattice \L is planar and self-dual.
Namely, if one places dual spins, one in each {\it face} of ${\cal L}$ 
including the face exterior to ${\cal L}$, and connects  dual spins with edges
crossing each of the edges of ${\cal L}$, one arrives at a lattice which
is precisely ${\cal L}$.

 Let  $Z_{\cal L}(q;K,L)$ be the partition 
function of the Potts model on ${\cal L}$.
Following the  remark after  (\ref{tmatrix}), one has  
  the  identity
\be
Z_{\cal L}(q;K,L) = q\> Z_N(q;K,L). \label{ghost}
\ee
It is then sufficient to show that  (\ref{duality}) holds
for $ Z_{\cal L}(q;K,L)$.
  
Generally, the Potts model 
possesses a duality relation 
for any planar lattice, or graph, with arbitrary edge-dependent interactions 
 \cite{wuwang}.
 Let $Z(K_{ij})$ be the partition function of the Potts model on a planar graph 
of $M$ sites with
edge-dependent interactions $K_{ij}$, and
$Z^{(D)}(K^*_{ij})$ the partition function  of the dual model.
Then,  the duality relation given in \cite{wuwang} can be written as
\be
{{q^{-M/2}Z(K_{ij})}\over { \prod_{\rm edges} \sqrt{e^{K_{ij}}-1}}} =
{{q^{-M_D/2}Z^{(D)}(K^*_{ij})}\over { \prod_{\rm edges} \sqrt{e^{K_{ij}^*}-1}}} ,
\label{duality1}
\ee
where $M_D$ is the number of sites of the dual graph and 
\be
(e^{K_{ij}}-1)(e^{K^*_{ij}}-1)=q.
\ee
Now the lattice \L is self-dual, namely,  $M=M_D=N+1$ and $Z=Z^{(D)}=Z_{\cal L}$.
The application of (\ref{duality1}) to \L now leads  to 
an expression which is precisely (\ref{duality})
with $Z_{\cal L}$ in place of $Z_N$.  
 The duality relation (\ref{duality}) now follows after introducing (\ref{ghost}).
 
\section{Partition function zeroes}
The zeroes of the partition function  can be computed from
\be
\l_+^N + \l_-^N +(q-2) (e^K-1)^N  =0, \label{zero}
\ee
at least numerically, for any given $N$. 
The analysis  is particularly simple if
  the three terms in (\ref{zero}) coalesce into two.
For the 
   $q=2$ Ising model, for example, one finds zeroes located at
 \be
\l_+ = e^{i(2n+1)\pi /N} \l_-, \hskip 1cm n=0,1,2,\cdots,N-1. \label{Isingzero}
\ee
This leads to the Yang-Lee circle \cite{yanglee} 
in the complex $e^{L}$ plane for $K\geq 0$.
 For  the zero-field Potts model, $L=0$,
 one has $\l_-=e^K-1$ and finds  from (\ref{zero})
the zeroes at 
\be
e^K+q-1 = 
(1-q)^{1/N} (e^K-1), \hskip 1cm n=0,1,2,\cdots,N-1. \label{Isingzero1}
\ee
 In the limit of $N\to \infty$, the partition function (\ref{partition1})
is dominated by the eigenvalue with the largest magnitude.  It follows that
the zeroes lie continuously on the  loci 
\begin{eqnarray}
&&|\l_+| = |\l_-|,
 \hskip 1.7cm {\rm in\>\>} |\l_+|\geq |e^K-1| \nonumber \\
&&|\l_\pm| = |e^K-1|, \hskip 1cm {\rm in\>\>} |\l_\pm|\geq |\l_\mp|,
\label{locus}
\end{eqnarray}
generalizing a previous result \cite{ms} for real $q$ and $L=0$.
Here, the loci (\ref{Isingzero1})
and (\ref{locus}) apply to real and complex $K, L, q$. 
Particularly, for $L=e^K=0$, 
$Z_N(q;K, L)$ gives the  ground state entropy
of the antiferromagnetic Potts model and the chromatic polynomial
(in $q$) of a ring.  In this case we have
$\l_+=q-1$, $\l_-=e^K-1=-1$, and   we find the zeroes   on the circle
 \be
|q-1|=1
\ee
in the complex $q$ plane.  This result 
has  previously been reported by Shrock and Tsai \cite{shrocktsai}.
 
For the Ising model the duality relation (\ref{duality}) 
provides some further  implications on the partition function zeroes.
  To conform with  usual notations, we consider  $N$
Ising spins on a ring with
nearest-neighbor interactions $K_{\rm I}$ and an external
magnetic field $L_{\rm I}\geq 0$.   The  Ising partition function is
\be
Z_{\rm Ising}(K_{\rm I}, L_{\rm I})
 =\sum_{\s_i=\pm 1} \prod_{i=1}^N\> 
[e^{K_{\rm I}\s_i\s_{i+1}}\>e^{L_{\rm I} \s_i}].
\label{isingp}
\ee
Using the identity $\s_i\s_{i+1}= 2\>\d_{\rm Kr}(\s_i,\s_{i+1})-1$
and $\s_i=2\>\d_{\rm Kr}(\s_i,1)-1$, we relate
$Z_{\rm Ising}$ to the  2-state Potts partition function via
\be
Z_{\rm Ising}(K_{\rm I}, L_{\rm I})
= e^{-N(L_{\rm I} +K_{\rm I})} Z_N(2;2K_{\rm I}, 2L_{\rm I}).\label{relation}
\ee
Thus, after substituting (\ref{relation}) into
(\ref{duality}) and some re-arrangement, we obtain
the following self-dual relation for the Ising model on ${\cal L}$,
\be
{{Z_{\rm Ising}(K_{\rm I},L_{\rm I})} \over {(\sinh 2K_{\rm I} \sinh 2L_{\rm I})^{N/4}}}
= {{Z_{\rm Ising}(K_{\rm I}^*,L_{\rm I}^*)} \over {(\sinh 2K_{\rm I}^* \sinh 2L_{\rm I}^*)^{N/4}}},
\label{isingd}
\ee
where the dual variables are 
\be
e^{-2K_{\rm I}^*} = \tanh L_{\rm I}, \hskip 1cm e^{-2L_{\rm I}^*} = \tanh K_{\rm I}.
\ee
We note the resemblance of  (\ref{isingd}) with the duality relation
of the Ising model
 formulated by Syozi \cite{syozi}.

From the Yang-Lee circle theorem \cite{yanglee}
we know that, for $K_{\rm I}\geq 0$, zeroes of 
$Z_{\rm Ising}(K_{\rm I}, L_{\rm I})$ lie on the unit circle
in the complex field $e^{-2L_{\rm I}}$ plane.  
Now, since we have $K_{\rm I}^*\geq 0$, the duality relation (\ref{isingd})
 implies that zeroes of  $Z_{\rm Ising}(K_{\rm I}, L_{\rm I})$ lie also
on the unit circle in the complex temperature
$\>\tanh K_{\rm I}\>$ plane, a result which  
holds for {\it any} 
$L_{\rm I}\geq 0$ and which is not readily recognized otherwise.
A consequence of this is that zeroes of
$Z_{\rm Ising}(K_{\rm I}, L_{\rm I})$ lie on the pure imaginary
axis  in the $e^{-2K_{\rm I}}$ plane, generalizing a result 
previously known only for $L_{\rm I}=0$ \cite{ms}.
Our circle theorem
contrasts with the square lattice result, 
valid only in the thermodynamic limit and in zero  field,
 which states that in the complex $\>\tanh K_{\rm I} \>$
plane zeroes of the partition function
$Z_{\rm Sq\> Ising}(K_{\rm I})$ lie 
 on two circles of radius $\sqrt 2$ and centered at $\pm 1$ \cite{fisher}.
 
\medskip
\noindent
{\it Note added}: After the completion of this manuscript, it has
been pointed out to me that results similar to those described 
here have been reported previously \cite{gu}.
The duality relation (\ref{duality}), which is derived in this manuscript
from a graphical consideration, was obtained in \cite{gu} 
algebraically by considering the transfer matrix. I would like to
thank A. Glumac for calling my attention to \cite{gu}.

\section{Acknowledgement}
Work has been supported in part by National Science Foundation
Grant DMR-9614170.  I would like to thank Professor R. Shrock for calling
my attention to Refs. \cite{ms,shrocktsai}.

\bigskip
\begin{center}
{\bf FIGURE CAPTION}
\end{center}
\medskip
\noindent
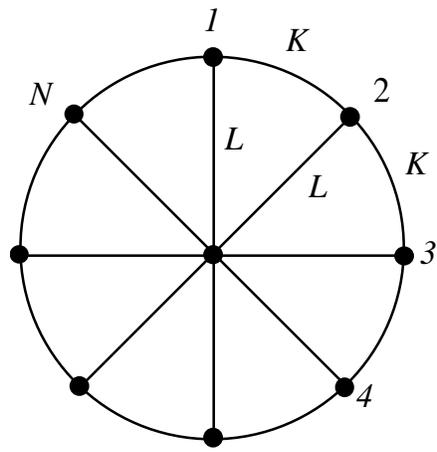
Fig. 1. The lattice ${\cal L}$ with $N+1$ sites.


\begin{thebibliography} {6}
\bibitem{eigen} The characteristic equation (\ref{char}) is factorized  by
subtracting the second row from  the third and each of the subsequent
rows,  followed by adding the third and all subsequent 
columns  to the second column.
\end{thebibliography}
\end{document}